\documentclass[iop]{emulateapj}

\usepackage{natbib}
\usepackage{graphicx}
\usepackage{multirow}

\slugcomment{{\sc }}
\usepackage[section]{placeins}

\newcommand{\lsim}{\mathrel{\hbox{\rlap{\lower.55ex\hbox{$\sim$}} \kern-.3em \raise.4ex \hbox{$<$}}}}

\def\Rtr{$R_{tr}$}
\def\mdot{\.{M}$_{out}$}
\def\mbh{$M_{BH}$}
\def\mic{~$\mu$m}

\slugcomment{}

\shorttitle{Disk, dust and jets in LLAGN}
\shortauthors{Mason et al.}

\begin{document}

\title{The role of the accretion disk, dust, and jets in the IR emission of low-luminosity AGN}

\author{R. E. Mason}
\affil{Gemini Observatory, Northern Operations Center, 670 N. A'ohoku Place, Hilo, HI 96720, USA}

\author{C. Ramos Almeida}
\affil{Instituto de Astrof\'\i sica de Canarias, C/V\'\i a L\'{a}ctea, s/n, E-38205, La Laguna, Tenerife, Spain}
\affil{Departamento de Astrof\'{i}sica, Universidad de La Laguna, E-38205, La Laguna, Tenerife, Spain}

\author{N. A. Levenson}
\affil{Gemini Observatory, Southern Operations Center, c/o AURA, Casilla 603, La Serena, Chile}

\author{R. Nemmen}
\affil{NASA Goddard Space Flight Center, Greenbelt, MD 20771, USA}

\and

\author{A. Alonso-Herrero}
\affil{Instituto de F\'{\i}sica de Cantabria, CSIC-UC, Avenida de los Castros s/n, 39005 Santander, Spain}
\affil{Augusto Gonz\'alez Linares Senior Research Fellow}

\begin{abstract}
We use recent high-resolution infrared (IR; 1 -- 20~$\mu$m) photometry to examine the origin of the IR emission in low-luminosity AGN (LLAGN). The data are compared with published model fits that describe the spectral energy distribution (SED) of LLAGN in terms of an advection-dominated accretion flow (ADAF), truncated thin accretion disk, and jet. The truncated disk in these models is usually not luminous enough to explain the observed IR emission, and in all cases its spectral shape is much narrower than the broad IR peaks in the data. Synchrotron radiation from the jet appears to be important in very radio-loud nuclei, but the detection of strong silicate emission features in many objects indicates that dust must also contribute. We investigate this point by fitting the IR SED of NGC~3998 using dusty torus and optically thin ($\tau_{\rm mid-IR} \sim 1$) dust shell models. While more detailed modeling is necessary, these initial results suggest that dust may account for the nuclear mid-IR emission of many LLAGN. 

\end{abstract}

\keywords{accretion, accretion disks --- galaxies: active --- galaxies: nuclei --- galaxies: individual (NGC 1052, NGC 3031, NGC 3998, NGC 4374, NGC 4486, NGC 4579, NGC 4594) --- infrared: galaxies}

\section{Introduction}
\label{intro}

Observational and theoretical studies of low-luminosity active galactic nuclei (LLAGN; AGN with $L_{bol} \lsim 10^{42} \rm \; erg \; s^{-1}$) suggest that these objects differ substantially from higher-luminosity AGN. At the low accretion rates typical of LLAGN, the inner part of the accretion disk is thought to take the form of a geometrically thick, optically thin advection-dominated accretion flow (ADAF), with the standard thin accretion disk limited to a region exterior to the ADAF. If LLAGN are analogous to X-ray binaries, the radius of the transition between the inner ADAF and outer disk should increase as the accretion rate declines \citep{Esin97,Narayan08}, and the thin disk may disappear altogether in quiescent objects. Strong winds and outflows are expected from ADAFs \citep[e.g.][]{Narayan95}, and indeed LLAGN are radio-loud compared to the general AGN population \citep{Ho02,Terashima03,Nagar05,Sikora07}.

The spectral energy distribution (SED) has been particularly useful in expanding our understanding of LLAGN. It has been instrumental, for instance, in the debate about whether LLAGN lack the ``big blue bump'' characteristic of the thin accretion disk in luminous AGN \citep[e.g.][]{Ho99,Maoz07,Eracleous10b}. The high-resolution data needed to separate the nuclear and host galaxy emission have been available for some time at radio, optical/ultraviolet and X-ray frequencies, but the first compilations of comparable infrared (IR) observations of substantial samples of LLAGN have only recently been published \citep[][hereafter M12]{Asmus11,Mason12}. The data show considerable variation among these objects. However, when considered in terms of Eddington ratio and radio loudness, they turn out to have some common characteristics. At low Eddington ratios (log $L/L_{Edd}< -4.6$), the IR-optical SEDs of radio-loud LLAGN (e.g. M87) are relatively flat (in $\nu L_{\nu}$) and may be dominated by synchrotron radiation. At higher $L/L_{Edd}$, the SEDs display a prominent peak in the mid-IR, and their IR-optical slopes tend to be within the range reported for higher-luminosity Seyferts.

The origin of the IR emission of LLAGN, and the process(es) responsible for the Seyfert-like IR peak in the higher-$L/L_{Edd}$ objects, are currently open questions. Several mechanisms may contribute: thermal emission from dust and/or the truncated accretion disk, and nonthermal emission from the jet and/or ADAF. The nuclear IR emission of radio-quiet Seyferts and quasars comes at least partly from dust in the torus \citep{Mor12}. However, some models predict that the torus should disappear in LLAGN, either because of a lack of material reaching the central engine from the host galaxy, or because of a disk wind ``switching off" at low accretion rates \citep{Elitzur06,Vollmer08}. The Seyfert-like IR-optical SEDs of some LLAGN, and the observation that these objects lie on the mid-IR/X-ray correlation established for higher-luminosity AGN, suggests that the torus could persist to fairly low luminosities \citep[][M12]{Asmus11}. The strong silicate emission features observed in a number of LLAGN, though, may indicate comparatively large amounts of optically thin dust, as expected if little material is reaching or being expelled from the nucleus.

While the accretion disk emission peaks in the UV in luminous AGN \citep{Sanders89}, that of the truncated disk is expected to shift to longer wavelengths. Consequently the IR emission of certain LLAGN has been suggested to arise in this component \citep{Ho08}, and several sets of ADAF+disk+jet models predict a prominent peak from the disk between $\lambda \sim$1 -- 10 \mic\ \citep[e.g.][]{Quataert99,Ptak04,Taam12}. Those models are barely constrained in the IR, however, because at the time of their publication, little high-resolution photometry was available.

\begin{figure}[]

\hspace{-0.4cm}
\includegraphics[scale=0.48]{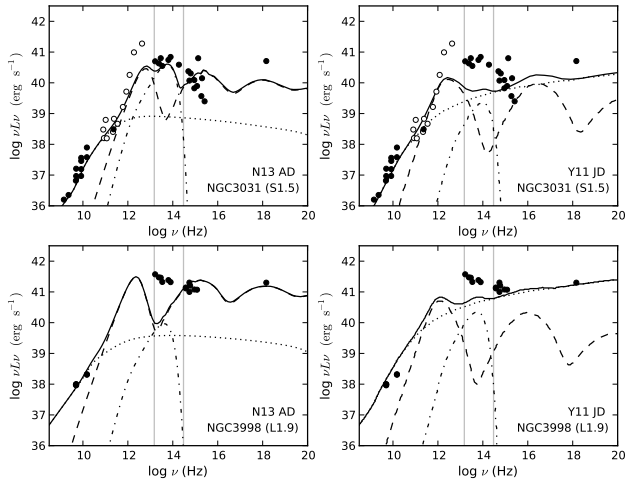}
\caption{ {\small Upper left: N13 ``ADAF-dominated'' (AD) model for
    NGC~3031. Upper right: Y11 ``jet-dominated'' (JD) model. The
    dashed line denotes the jet, the dotted line the ADAF, the
    dot-dashed line the truncated disk, and the solid line their
    sum. Nuclear data are shown as solid circles, while large-aperture
    ($>1$\arcsec) points are indicated by open symbols. The gray lines
    in each figure indicate 1~$\mu$m and 20~$\mu$m. Lower panels: As
    for upper panels, for NGC~3998.}} 
\label{f1}
\end{figure}

Synchrotron emission from a jet may also play a r\^{o}le in some objects. Mid-IR interferometric observations of the radio-loud LLAGN, Centaurus A, show two components: a resolved, disk-like structure and an unresolved core \citep{Meisenheimer07}. While the disk component, which accounts for 20-40\% of the 8--13~$\mu$m parsec-scale flux, is interpreted as dust heated by the AGN, the de-reddened core is well fit by a synchrotron model that turns over in the mid-IR.

Many LLAGN have now been modeled in terms of a jet, ADAF and truncated thin disk. The models often provide a good fit to the high-resolution radio, optical, UV and X-ray data and have been used to estimate quantities such as mass accretion rates in LLAGN (Nemmen et al. 2013, submitted; hereafter N13)\footnotemark{}. As a first step towards understanding the IR emission of these objects, in this Letter we compare a representative set of these existing ADAF+disk+jet model fits with SEDs that include the newly-available high-resolution IR data. 

\footnotetext[1]{The original version of N13 is available at http://adsabs.harvard.edu/abs/2011arXiv1112.4640N. N13 differs in that (1) only 12 objects with well-sampled SEDs are chosen for detailed study; (2) jet-dominated models (\S\ref{models}) are not considered for objects with X-ray photon index $\lsim2$; (3) some model fits have changed, e.g. a luminous thin disk component is no longer included for NGC~4374 as the SED shows little evidence for its presence.}
\section{Model comparisons}
\label{models}

\begin{figure}

\hspace{-0.4cm}
\includegraphics[scale=0.48]{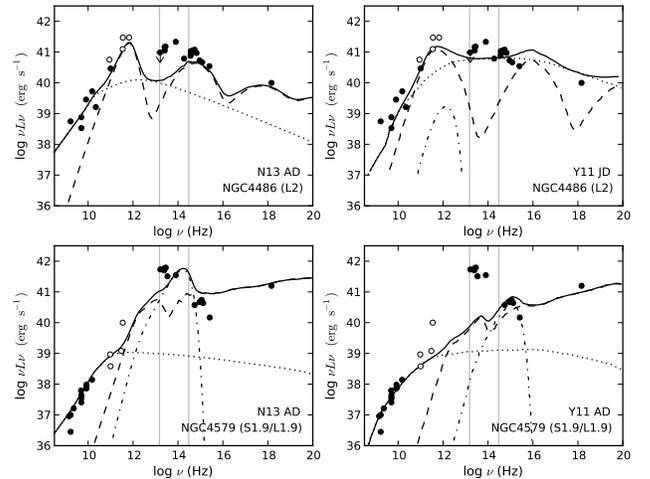}
\caption{ {\small As for Figure \ref{f1}, for NGC~4486 and NGC~4579. The truncated disk spectrum for NGC~4486 is not given by N13 as its luminosity is negligible. Y11 and N13 do not present JD models for NGC~4579, so both AD models are shown here.}}
\label{f2}
\end{figure}

\begin{figure}

\hspace{-0.4cm}
\includegraphics[scale=0.48]{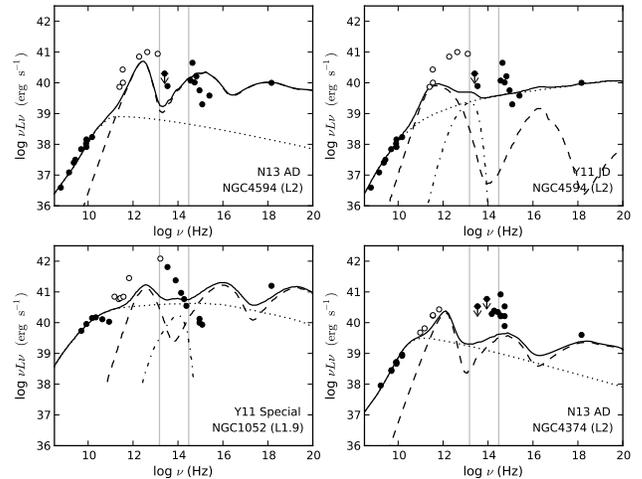}
\caption{ {\small As for Figure \ref{f1}, for NGC~4594, plus two objects, NGC~1052 and NGC~4374, for which only a single ADAF+disk+jet model has been published. The truncated disk spectrum for NGC~4374 and NGC~4594 is not given by N13 as its luminosity is negligible. The Y11 model for NGC~1052 is neither AD nor JD.}}
\label{f3}
\end{figure}

Of the 22 LLAGN in M12, ADAF+disk+jet models have been published for 14 \citep[][]{Lasota96,Quataert99,Gammie99,Yuan02,Ptak04,Nemmen06}. We select seven for further study (Figures \ref{f1} -- \ref{f3}). The remaining galaxies' SEDs may contain a strong contribution from dust heated by nuclear star formation \citep[NGC~1097; ][]{Mason07}, are strongly affected by foreground extinction \citep[NGC~4261; ][]{Eracleous10b}, or are very sparsely sampled and/or may contain significant host galaxy emission. NGC~4258 is discussed by \citet{Wu13}. 

The most extensive and uniform sets of model fits are those of N13 and \citet[][hereafter Y11]{Yu11}, so we use those for this comparison \citep[see][N13 for details of the underlying models]{Yuan05}. The ADAF lies within the transition/truncation radius, \Rtr, outside which is found the truncated thin disk (see Figure 1 of N13). The ADAF spectrum depends on the black hole mass \mbh, accretion rate \.M, and other factors such as the viscosity parameter $\alpha$ and the fraction of the turbulent dissipation heating the electrons $\delta$. The accretion rate varies with radius, with \mdot\ denoting the accretion rate at \Rtr. The mass loss rate in the jet, \.M$_{jet}$, is required to be a reasonable fraction of the accretion rate in the ADAF's inner region (Y11), or not to exceed \mdot\ (N13).

The thin disk emits locally as a blackbody and is described by \mbh\ (estimated from stellar velocity dispersions or gas kinematics) and the inclination angle $i$ (often constrained by radio jet observations, modelling of the broad H$\alpha$ line or other measurements), along with \Rtr\ and \mdot. \Rtr\ and \mdot\ are correlated, but the main effect of \Rtr\ is to determine the spectral cutoff of the disk emission (smaller \Rtr\ implying a higher-frequency cutoff). \Rtr\ has little effect on the ADAF SED, as most of the ADAF emission comes from the inner region of the flow.

Y11 make initial estimates of $R_{tr}$ and \.M$_{out}$, then combine these values with the other parameters ($M_{BH}$, $\alpha$, $\delta$, etc.), which are fixed, to derive the ADAF solution and SED. The disk spectrum is calculated using these initial values of $R_{tr}$ and \.M$_{out}$, the jet SED computed, and the sum of the emission of these components compared to the observed SED. The input parameters are adjusted until a model is obtained that is roughly consistent with the data. The radio and/or X-ray data constrain \.M$_{out}$, which is correlated with \Rtr.

N13 first fit a jet model to the radio data, then fit the truncated disk to the optical photometry to obtain $R_{tr}$ and \.M$_{out}$. The X-ray data are then used to refine the values of $R_{tr}$ and \.M$_{out}$, estimate 
$\delta$, and derive the ADAF model. The emission from the various components is then summed and compared to the overall broad-band SED. Both N13 and Y11 refer to two families of models: ``ADAF-dominated"  (AD) and ``jet-dominated" (JD), depending on which is the dominant contributor to the X-ray emission. The models are generally fit only to radio, optical, UV and X-ray data; existing low-resolution IR points are treated as upper limits.

Figures \ref{f1} - \ref{f3} show the comparison between the observed SEDs \citep[from M12 and references therein, plus Herschel far-IR data for NGC~3031 from][]{Bendo12} and published ADAF+disk+jet models for these objects. The M12 SEDs rely heavily on those of \citet{Eracleous10b}, with the addition of high-resolution IR data obtained from (near-)diffraction-limited MIR, adaptive optics and HST imaging. Outside the IR region there is a large degree of overlap between the M12 SEDs and those fitted by N13 and Y11, but they are not identical. The N13 optical/UV data use the X-ray-based extinction corrections of \citet{Eracleous10b} while those of M12 and Y11 are not corrected for extinction, and the X-ray (and some other) points chosen by each set of authors are not the same in all cases. In view of the heterogeneity of the data collections and extinction corrections, our focus here is simply to compare the model fits and SEDs in the IR region, rather than to re-examine in detail the model results over all frequencies. The reader is encouraged to consult Y11 and N13 to see the model fits to the original SEDs.

Y11 give their preferred (AD or JD) model for each nucleus, whereas N13 show both AD and JD versions unless one results in physically unreasonable parameters. Where possible, then, we show one AD and one JD model for each object. The comparison between the models and nuclear IR data shows that:

1) The models are usually consistent with the nuclear IR data, in that they do not predict stronger emission than is observed.  NGC~4579 may be an exception. Although its SED lacks the high-resolution 1 -- 3~$\mu$m data needed to make a definitive statement, the truncated disk in the N13 AD model (Fig. \ref{f2}) will produce too much NIR emission unless the spectral shape of this object is highly unusual. Similarly, \citet{Wu13} fit an AD model to NGC~4258 and find that the disk is over-luminous in the NIR. Conversely, the models never account for {\em all} of the  emission over the whole 1 -- 20 $\mu$m range. There is always room for another component to contribute, particularly in the mid-IR.

2) The spectral shape of the truncated disk emission does not bear a good resemblance to the IR SEDs, tending to be much narrower than the broad IR peaks in the observations. It is conceivable that the disk and ADAF emission could combine so as to reproduce the broad IR peak in the SEDs, but this is unlikely to happen in every object. In any case the luminosity of the truncated disk component in most of the models is much lower than the observed IR luminosity.

3) The ADAF synchrotron emission, which peaks at $\nu \sim 10^{12}$ Hz, is accompanied by a Compton scattering bump at $\nu \sim 10^{15} - 10^{16}$ Hz. These two components usually bracket the nuclear IR emission.

4) The contribution of the combined ADAF+disk+jet to the observed IR emission is rather model-dependent. In NGC~3031 the Y11 JD model accounts for very little of the IR luminosity, whereas the N13 AD model predicts that most of the emission at $\lambda \sim 10 \; \mu$m comes from the truncated disk. The AD models tend to predict an SED whose luminosity increases sharply towards optical wavelengths, while the JD models are flatter. However, different models from the same family can also give very different results. The N13 AD model for NGC~4579 predicts strong NIR emission from the truncated disk, but the disk emission in the Y11 AD model peaks at optical wavelengths. In both models \Rtr\ is consistent with constraints from the broad H$\alpha$ line \citep{Barth01}.

\section{Discussion/Conclusions}
\label{discuss}

Perhaps the most notable aspect of the comparison between the ADAF+disk+jet model fits and high-resolution SEDs is the presence of an IR ``excess'' above the predictions of the models. Except in one or two cases, the truncated disk is not luminous enough to account for the IR emission even over a limited range of wavelengths. The fits presented in Y11 and N13 are not unique, and it may be possible to identify ADAF models that fit the radio and X-ray emission while allowing a more luminous disk component. However, the spectral shape of the disk emission remains a poor match to the overall 1 -- 20 $\mu$m data. We conclude that the truncated accretion disk is unlikely to be responsible for the IR ``bump'' in LLAGN, contrary to some recent suggestions \citep[e.g. ][]{Ho08,Taam12}. 

The contribution of the ADAF can be significant around $\lambda \sim 1 \; \mu$m, but it drops off sharply towards longer wavelengths in almost all of the models. The peak frequency of the ADAF synchrotron emission depends on \mbh\ and $\dot{M}$
\citep{Mahadevan97}, and occurs at $\nu_{\rm peak} \sim 10^{12}$ Hz for these LLAGN. The characteristic Compton boost parameter is such that the first peak due to the upscattering of seed synchrotron photons is typically around $10^{15}-10^{16}$ Hz. The synchrotron and Compton peaks bracket the 1-20 $\mu$m ($\sim 10^{13}-10^{14}$ Hz) observations, disfavoring an ADAF origin for the nuclear IR emission.

The models do include synchrotron radiation from a jet, but the jet generally does not account for all of the IR emission. 
In the Y11 model for the radio galaxy NGC~1052, the jet dominates the emission at $\lambda \sim 1 \; \mu$m but is negligible at longer IR wavelengths.  However, \citet{Fernandez-Ontiveros12} point out that the optical -- IR emission of that object resembles a power-law, and are able to fit the entire radio -- UV SED (not corrected for extinction) with a simple self-absorbed synchrotron spectrum. This may suggest that the nuclear IR data of radio-loud objects should be included when fitting their SEDs with the ADAF+disk+jet models, potentially leading to very different derived model parameters. 

\begin{figure}
\includegraphics[scale=0.45]{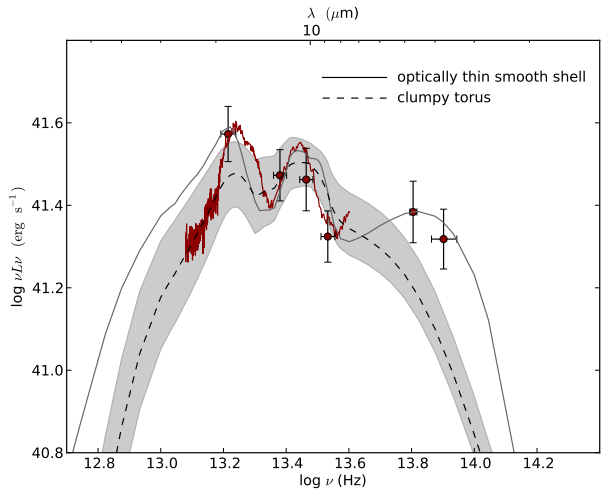}
\caption{ {\small Clumpy torus and optically thin dust model fits for NGC~3998. The solid gray line indicates the best-fitting optically thin dust model, determined by $\chi^2$ minimisation. The dashed line shows the clumpy torus model fit produced using the maximum a-posteriori values that represent the best fit to the data, while the shaded region denotes the range of models compatible with a 68\% confidence interval. The filled circles show the high-resolution photometric data points, while the red line is the Spitzer spectrum of this object ($\times$ 0.6). }}
\label{f4}
\end{figure}

A number of the LLAGN studied here, including NGC~1052, have remarkably strong silicate emission features in their spectra (M12). Many also have IR SED slopes within the range spanned by ``conventional'' Seyferts \citep[e.g.][]{RamosAlmeida09,RamosAlmeida11,Prieto10}. The silicate features are an unambiguous indicator of dust, and their strength in these LLAGN suggests that they arise in dust heated by the AGN rather than in dust associated with circumstellar shells in the host galaxy \citep[c.f.][]{Bressan06}. These observations imply that dust must contribute at some level to the nuclear IR emission of LLAGN. In Figure \ref{f4} we show two dusty model fits to the IR SED of NGC~3998: optically thin dust emission produced with the DUSTY radiative transfer code \citep{Ivezic99}, and 
a clumpy torus spectrum obtained using the models of \citet{Nenkova08a} and the Bayesclumpy fitting tool \citep{AsensioRamos09,AsensioRamos13}.
Clumpy torus models have been successful in fitting the SEDs of numerous Seyfert galaxies \citep[e.g.][]{Mason06,Honig10b,Alonso-Herrero11}, while the presence of the strong silicate emission features in this object \citep[][M12]{Sturm05} suggests the use of the optically thin dust model.

In the torus model in Figure \ref{f4}, the bolometric luminosity required to account for the IR emission of NGC~3998, $L_{bol} = 1 \times 10^{42} \rm \; erg \; s^{-1}$, is comparable to that estimated from the X-rays \citep[$L_{bol} \sim 3 \times 10^{42} \rm \; erg \; s^{-1}$, assuming the bolometric correction factor of][]{Ho09}. The $N_H \sim 8 \times 10^{22} \rm \; cm^{-2}$ implied by the visual extinction through the clumpy torus, $A_{V} = 41 \pm 4$ mag, is quite close to the observed $N_H \sim 2.3^{+3.2}_{-1.6} \times 10^{22} \rm \; cm^{-2}$ \citep{Gonzalez-Martin09b}, and the escape probability of an AGN-produced optical photon, $P_{esc} = 0.19$, is within the range spanned by the type 1 objects in \citet{Alonso-Herrero11}. The  torus models can produce a wide range of SED shapes and the photometry alone does not restrict the parameter space to a high degree. For that reason, the Spitzer spectrum was included in the fits. Although the angular resolution of the spectroscopy is a factor of $\sim$10 lower than that of the ground-based photometry, its spectral shape matches the photometric data well. With the constraints imposed by the spectroscopy, the torus model does predict silicate features in emission, although not as strongly as observed. 

The torus appears under-luminous at $\lambda < 5 \; \mu$m. Given the uncertainty in the luminosity of the ADAF+disk+jet in the near-IR (\S\ref{models}), it is not yet clear whether the torus model is deficient in hot dust emission, or whether the observed emission at these wavelengths contains a contribution from sources other than dust. In the ADAF+disk+jet models considered for NGC~3031, the largest contribution to the observed IR emission arises in the Y11 model around 3 -- 4 $\mu$m, where the model would account for $\sim30$\% of the observed luminosity. At longer wavelengths the ADAF+disk+jet contribution would be within the photometric errors. We have subtracted the Y11 model from the data and refit the residual emission, finding only minor changes to the derived parameters ($L_{bol} = 1 \times 10^{42} \rm \; erg \; s^{-1}$, $A_{V} = 38 \pm 5$ mag, $P_{esc} = 0.30$).

The basic features of the optically thin dust model are described in \citet{Sirocky08}. Briefly, the dust is distributed in a spherical shell around a broken power law AGN spectrum, from a minimum radius corresponding to the dust sublimation temperature (1500 K). Dust that is optically thin in the mid-IR naturally produces silicate features in emission, so the Spitzer IRS spectrum was not included in these fits. The best-fitting model (Figure \ref{f4}) provides a reasonable description of the SED of NGC~3998, and the strength of the silicate emission features is comparable to those observed in the Spitzer spectrum. More sophisticated treatment of grain chemistry/size/structure and/or radiative transfer effects may be necessary to improve the match between the observed and modelled feature profiles \citep[e.g.][]{Nikutta09,Smith10}. The V-band optical depth of the model, $\tau_{V}=33$, means that the AGN is heavily obscured in the optical. This is not consistent with the detection of broad $H\alpha$ emission in this object \citep{Devereux11}. A toroidal dust distribution, not considered in this simple model, would allow direct views of the broad line region without significantly affecting the reprocessed spectrum.

More in-depth modeling of this and other nuclei will certainly be necessary for a full understanding of the IR emission mechanisms of LLAGN. If confirmed, optically thin dust emission would be consistent with a ``disappearing torus'' scenario in which clouds forming in the accretion disk wind of a low-luminosity AGN do not become sufficiently optically thick to remain coherent and form an obscuring torus \citep{Elitzur06}. Additional high-resolution near-IR imaging and mid-IR spectroscopy would allow firmer constraints to be placed on the dusty models, and IR polarimetry
\citep[with MMT-POL, for example;][]{Packham12} may help disentangle the contributions of dust and jet emission to the IR luminosity of radio-loud LLAGN. However, on the basis of the IR ``excess" above the ADAF+disk+jet model fits examined here, the mismatch between the shape of the observed IR SED and that of the ADAF+disk+jet models, and the ability of dusty models to account for the IR SED, silicate emission and other characteristics of NGC~3998, we suggest that dust contributes strongly to the emission of many low-luminosity AGN between $\sim$ 1 -- 30 $\mu$m. 

\acknowledgments
We thank the referee for a useful report that helped improve this work. Supported by the Gemini Observatory, operated by the Association of Universities for Research in Astronomy, Inc., on behalf of the international Gemini partnership of Argentina, 
Australia, Brazil, Canada, Chile, and the USA. A.A.-H. acknowledges support from the Spanish
Plan Nacional de Astronom\'{\i}a y Astrof\'{\i}sica under grant AYA2009-05705-E, CRA from PN-AYA2010-21887-C04.04.


\end{document}